\documentstyle[aps,multicol,epsf]{revtex}
\catcode`\@=11

\def\eqbegin         {  \begin{eqnarray}  }
\def\eqend           {  \end{eqnarray}  }
\def\beq{\begin{equation}}
\def\eeq{\end{equation}}

\def\del          { \partial }

\def\Z{{\bf Z}}

\def\hs_2{\hspace{2mm}}
\def\hs_3{\hspace{3mm}}

\title{Spin-singlet hierarchy in the fractional quantum Hall effect }
\begin{document}
\maketitle
\begin{center}
{Kazusumi Ino}\\
{\em Nomura Research Institute,} {\em Hongo 2-2-9,  Bunkyo-ku,  Tokyo,
113-0033, Japan}
\end{center}

\begin{abstract}
 We show that   the so-called permanent quantum Hall states are
 formed by the integer quantum Hall effects on
the Haldane-Rezayi quantum Hall state.
Novel conformal field theory description along with this picture
is deduced.
The odd denominator plateaux observed around $\nu=\frac{5}{2}$
 are  the permanent states if the $\nu=\frac{5}{2}$
 plateau is the Haldane-Rezayi state.
We point out that there is no such hierarchy on
other candidate states for $\nu=\frac{5}{2}$.
We  propose  experiments to test our prediction.
\end{abstract}

\pagenumbering{arabic}
\begin{multicols}{2}
In 1987, Willet et al discovered a quantum Hall effect  at
an even-denominator filling fraction  $\nu=5/2$ \cite{willet}.
This is so far
the only even-denominator Fractional Quantum Hall Effect (FQHE)
found in a single-layer system.
The tilted field experiments suggest that the plateau
 is a spin-unpolarized singlet state \cite{eisen}.
This motivated Haldane and Rezayi to propose a variational
ansatz for the spin-singlet ground state,
the Haldane-Rezayi (HR) state \cite{halrez}.
Although it has satisfactory features as a candidate for
the $\nu=5/2$ plateau, it was shown that
the hollow-core Hamiltionian
with which the HR state is the exact ground state does not
reproduce the realistic Hamiltonian even in the second Landau level
\cite{YGM2}.
Recent numerical studies \cite{morf} show
that the ground state may be a spin-polarized
pfaffian-like  state \cite{moore}.
If the state is spin-polarized, however,
new explanation for the results of tilted
field experiments is required.  Some suggestions, such as the effect of
thickness of real sample have been  made.
However, the discussions are not satisfactory
yet and the final answer remains to be seen.

The  enigmatic $\nu=5/2$ plateau is a FQHE at the  second Landau level.
In the FQHEs in the lowest Landau level,
the observed plateaux  are well described by the
Jain's hierarchy \cite{jain}.  The key of Jain's idea is
the non-perturbative object called
{\it composite fermion}.  It is   formed by the electron and the vortex.
At $\nu=1/2$, the external magnetic field is completely
screened by the formation of composite fermions and
 they form an unusual metallic state ;
  a Fermi liquid of composite fermions \cite{HLR}.
The Jain's picture is that the FQHEs
are the {\it integer} QHEs in the vicinity of the filling
fraction $\nu=1/2$.
The Landau levels emerging here are those for composite fermion.
This scheme predicts most of the observed FQH plateaux around
 $\nu=1/2$.

It is worth while asking whether similar
scheme holds in the second Landau level.
It seems natural to consider that the
$\nu=5/2$ plateau might have the same role in the FQHE at
 the second Landau level as the  $\nu=1/2$ state
 does in the FQHE   at the lowest Landau level.
Suppose that the Fermi surface of the Fermi liquid  of
composite fermions has an instability to
form a pair condensate.
Even in this case,  the Landau levels  are formed by adding
magnetic field  and  QH plateaux would appear.
These plateaux are not expected to be the ordinary FQH state :
they should inherit the feature
of the state at $\nu=5/2$ such as pairing of
composite fermions.
It suggests that
the odd-denominator plateaux around $\nu=5/2$
($\nu=7/3,8/3 ...)$ observed in Ref.\cite{willet}
  would also be a part of the enigma.

In this paper, we will give evidences
 that this will be the case.
We show  that  a class of spin-singlet QH states,
namely the so-called permanent QH states
(which are variants of the HR state first proposed in Ref.\cite{YGM})
are formed  by integer QH effects around the
HR state.   This observation is supported by the existence of
conformal field theory (CFT) description of novel kind which we
deduce below.
Based on this novel description, we develop a
hierarchical scheme of the permanent QH states.
It yields a candidate comprehensive scheme for
 FQH plateaux $ \nu = 7/3,8/3 ...$ etc
around $\nu=5/2$.
We also point out that
such a construction only exists for the HR state i.e.
 no such hierarchy for Pfaffian or other candidate states for
 the $\nu=5/2$ plateau exists.
If we believe in the validity of the idea of hierarchy
around $\nu=5/2$,
the existence of the construction gives an evidence that
the $\nu = 5/2$ plateau is the HR state.

First we note that Bonesteel
recently showed that composite fermions without additional degrees of freedom
do not form a pair condensate by themselves \cite{bonesteel}
based on the celebrated theory of Ref. \cite{HLR}.
Thus the essential ingredient  which should be studied is  the degrees of
freedom which drive composite fermions to form a pair.

Let us consider the relation between the
simplest permanent state and the HR state.
The wave function of the
permanent state at $\nu=\frac{1}{q}=\frac{1}{p+1}$ ($q=p+1$,$p$: even integer)
has a simple form,
\\
$ \Psi_{\rm per} = $
\eqbegin
{\rm per}\left( \frac{1}{z^{\uparrow}_i-z^{\downarrow}_j}
\right) \prod (z_i-z_j)^{q}
{\rm exp}\left(-\frac{1}{4}\sum_i^{N} |z|^2 \right),
\label{per}
\eqend
where $N$  is the number of electrons.
The HR state at $\nu=\frac{1}{p}$ ($p=2$ for $\nu=5/2=2+1/2$)
 has a more involved form
(we  omit the exponential factor hereafter),
\\
$\Psi_{\rm HR} = $
\eqbegin
{\rm per}
\biggl( \frac{1}{z^{\uparrow}_{i}-z^{\downarrow}_{j}}\biggr)
\prod(z^{\uparrow}_i-z^{\downarrow}_j)^{q-2}
\prod(z^{\uparrow}_i-z^{\uparrow}_j)^{q}
\prod(z^{\downarrow}_i-z^{\downarrow}_j)^{q}.
\eqend
From these forms, it seems that the HR state
can be realized by some additional degrees of freedom and
structure on $\Psi_{\rm per}$.
However the HR state also has the following form:
\eqbegin
\Psi_{\rm HR} = {\rm det}\left( \frac{1}{(z^{\uparrow}_i-z^{\downarrow}_j)^2}
\right) \prod (z_i-z_j)^p
\label{HRwave}
\eqend
It is known that
the permanent factor in (\ref{per}) and the determinant in (\ref{HRwave})
 have a simple relation \cite{halrez}
\eqbegin
{\rm det}\left( \frac{1}{(z^{\uparrow}_i-z^{\downarrow}_j)^2}\right) =
{\rm per}\left( \frac{1}{z^{\uparrow}_i-z^{\downarrow}_j}\right)
{\rm det}\left( \frac{1}{z^{\uparrow}_i-z^{\downarrow}_j}\right).
\label{perdet}
\eqend
This can be rewritten as
\eqbegin
{\rm per}\left( \frac{1}{z^{\uparrow}_i-z^{\downarrow}_j}\right)
=
{\rm det}\left( \frac{1}{(z^{\uparrow}_i-z^{\downarrow}_j)^2}\right)
{\rm det}\left( \frac{1}{z^{\uparrow}_i-z^{\downarrow}_j}\right)^{-1}.
\eqend
 This shows that the wave function of the permanent state can be
 written as
\eqbegin
 \Psi_{\rm per} &=&
\Biggl[ {\rm det}\left( \frac{1}{z^{\uparrow}_i-z^{\downarrow}_j}\right)^{-1}
\prod (z_i-z_j)\Biggr]
\Psi_{\rm HR}.
\label{factorization}
\eqend
This factorization is  just the reminiscent of the one for the Laughlin wave
 function in the Jain's  paper \cite{jain}.
The Langhlin factor in (\ref{factorization})  implies the formation of
the filled Landau level  i.e. the $\nu=1$ integer QH effect.
The factor of the inverse of determinant is a novel one and requires
an interpretation.

In general, the composite fermions in the Landau level
repulsively interacts each other by the Coulomb interaction.
This gives rise to the Laughlin factor.
When the composite fermions are paired, it would  also
have an effect on the strength of pairing.
As ${\rm det}(\frac{1}{z^{\uparrow}_i-z^{\downarrow}_j})$ has poles of
order one, ${\rm det}(\frac{1}{z^{\uparrow}_i-z^{\downarrow}_j})^{-1}$
has zeroes of order one for each pair.
Thus it will reduce  the strength of pairing by order one.
The reduction is physically
interpreted as  the effect of
the formation of the filled Landau level on the pairing.

Thus the permanent states
 can be seen as the integer QHEs around the HR state.
Extension to multiple filled Landau levels is straight forward.

As it turns out, this observation  realizes in the CFT description.
As  observed in \cite{moore}, the pairing part of the permanent state can be
written as the correlator of the $c=-1$
bosonic ghosts $\beta$-$\gamma$ \cite{FMS},
 whose conformal dimensions $\Delta$ are $1/2$
(we also denote them as $\beta^{\alpha},
\hspace{2.5mm}\alpha =\uparrow,\downarrow$) :
\eqbegin
{\rm per}\left( \frac{1}{z^{\uparrow}_i-z^{\downarrow}_j}
\right) = \langle \prod_{i=1}^{N/2}
\beta(z_i^{\uparrow})\gamma(z_i^{\downarrow})
\rangle .
\eqend
In terms of these fields and a chiral boson $\varphi$
($\langle \varphi(z_1)\varphi(z_2) \rangle=-{\rm log}(z_1-z_2)$),
the electrons in the permanent state are represented by
$\beta^{\alpha} e^{i\sqrt{q}\varphi}$.
There is an ingenious bosonization for the $\beta$-$\gamma$ system \cite{FMS},
in which $\beta$ and $\gamma$ are bosonized into
$c =-2$ $\xi$-$\eta$ system and a chiral boson $\phi$
with negative signature
as follows:
\eqbegin
\beta = \del \xi e^{\phi},\hspace{5mm}
\gamma = \eta e^{-\phi}.
\label{boso}
\eqend
(\ref{boso}) reproduces the formula (\ref{perdet}).
Note that the zero mode of $\xi$ does not  appear in this bosonization.
As in \cite{FMS}, the twisted sector is usually created by
the spin field $\Sigma=e^{\phi/2}$ of $\Delta =-1/8$.
By this field, the elementary quasiholes in the permanent state is
given by $\Sigma e^{\frac{1}{2\sqrt{q}}\varphi}.$
As noted in \cite{RR},  the twisted
 sector is infinitely degenerated, due to the appearance of
 the zero mode of $\beta^{\alpha}$.

Although one can construct the bulk wave functions by this system,
it does not necessarily imply that it is the right description
of the permanent state.
One way to check it is to consider the consistency of
the edge theory which governs the low energy physics
at the thermodynamic limit $N \rightarrow \infty$.
The edge excitations are generated
by the $\beta$-$\gamma$ system and $i\sqrt{\nu} \del \varphi$.
The excitations
on each sector contribute to the partition function
through the Virasoro characters :
\eqbegin
\chi_{1}& =&\frac{1}{2}\left(\frac{\eta(t)}{\vartheta_3(t)}
+\frac{\eta(t)}{\vartheta_2(t)}\right),\\
\chi_{\beta}&=&
 \frac{1}{2}\left(\frac{\eta(t)}{\vartheta_3(t)}
 -\frac{\eta(t)}{\vartheta_2(t)}\right),\\
\chi_{\Sigma}&=& \frac{\eta(t)}{\vartheta_4(t)},
\label{c=-1chara}
\eqend
where $t$ is $-\frac{1}{\tau}$ ($\tau$ is the modular parameter).
In the twisted sector the degenerated ground states
are summed up and regularized \cite{RR,gurus}.
The characters for the U(1) sector are given by
$\chi_{ r/q}(t) =  \frac{1}{\sqrt{q}}
\frac{\vartheta_3(\frac{r}{q}|\frac{t}{q})}{\eta(t)}$,
$\chi^{\vee}_{r/q}(t)
=\frac{1}{\sqrt{q}}  \frac{\vartheta_2(\frac{r}{q}|\frac{t}{q})}{\eta(t)}$
and $\chi_{(r+1/2)/q}(t)
=\frac{1}{\sqrt{q}}
 \frac{\vartheta_4(\frac{r}{q}|\frac{t}{q})}{\eta(t)}.$
To check the consistency,
we consider the edge theory for the state on the cylinder.
The edge states on two edges are divided into
finite sectors up to the electrons,
and the partition function has a corresponding factorization.
Eventually we end up with three terms:
\eqbegin
V=\frac{1}{2q}\sum_{r=1}^{q}\left
|\frac{\vartheta_3(\frac{r}{q}|\frac{t}{q})}{\vartheta_3(t)}\right|^2,
\\
W=\frac{1}{2q}\sum_{r=1}^{q}\left|
\frac{\vartheta_2(\frac{r}{q}|\frac{t}{q})}{\vartheta_2(t)}\right|^2,
\\
X=\frac{1}{q}\sum_{r=1}^{q}\left|\frac{\vartheta_4(\frac{r}{q}|\frac{t}{q})}{\vartheta_4(t)}\right|^2.
\eqend
The problem here is that the multiplicities
of the last two characters $V, W$ and $X$ do not match, thus  do not
satisfy  a necessary condition for
the {\it modular invariance} \cite{cappelli}.
To achieve the matching,
we need additional zero modes which give additional degeneracy
to $V$ and $W$. One expect that the zero mode $\xi$ may play a role like in
 the picture changing operation of superstring theories \cite{FMS}.
However, naive addition of $\xi$ will add the factor 2
to  $V,W$ and $X$ equally.

We note  that these factors $2$ from $\xi$  arises
since we create the twisted sector  by $e^{\phi/2}$.
Actually this is not the only possibility;
one can instead twist the $c=-2$ part.
Twisiting  the $c=-2$ part will maximally
extend the $\xi$-$\eta$ system to the symplectic fermions' \cite{kausch}.
We thus should deal with the symplectic fermions $\theta^{\uparrow},
\theta^{\downarrow}$. Now $\beta$ and $\gamma$ are bosonized into
\eqbegin
\beta=\del\theta^{\uparrow}e^{ \phi}, \hspace{5mm}
\gamma=\del\theta^{\downarrow}e^{-\phi}.
\label{bonzo}
\eqend
The twisted sector of these fields is created by $\sigma$,
the spin field for the symplectic fermions.
The operator product of the spin fields
 is not closed within the $\beta$-$\gamma$ system,
and the zero modes of $\theta$
naturally appear for consistency \cite{kausch}.
Thus this system is  the extended $\beta$-$\gamma$ system.
By virtue of the twisting we take,
 the zero modes of $\theta$ do not appear in the twisted sector,
 just leading  to the   multiplicities required above.
The partition function $Z=2V+2W+X$ is modular invariant.
The five sectors of the partition function correspond to
the fields $1,\hspace{2mm}\beta^{\alpha},\hspace{2mm}\sigma,
\hspace{2mm}\widetilde{I}(=\theta\bar{\theta})$ and
$\beta^{\alpha}\widetilde{I}$
($\theta\bar{\theta} =\frac{1}{2}(\theta^{\uparrow}\theta^{\downarrow}-
\theta^{\downarrow}\theta^{\uparrow}$).)
Accordingly the permanent state on the torus have
$5q$ degeneracy. The torus wave functions are obtained as
 the conformal blocks  using
the torus version of the bosonization (\ref{bonzo}).

The conformal field theory we end up with  supports  our observation that
the permanent states are formed by
the integer QH effects around  the
HR state. First the electron fields are represented by
the same degrees of freedom (symplectic fermions)
of the electron fields in the HR state \cite{milo,gurarie,ino},
together with  additional chiral boson $\phi$ which
 is regarded as describing  the
effect of the formation of the filled Laudau level on the pairing.
The quasiholes are also constructed from $\sigma$ which comes
from the  quasiholes of the HR state.
The appearance of the zero modes of $\theta$, thus the presence
of the logarithmic field $\widetilde{I}$ is also the reminiscent
of the HR state. This implies that, in the formation
of the permanent state, these zero modes
from the d-wave pairing in the HR state survive.

As shown in \cite{ino},  the edge state of the HR state
has an instability to generate a flux quantized to half integer.
The permanent state does not inherit this feature.
The flux which couples to the spin is ill-defined in this case,
 so we can consider the Aharanov-Bohm flux.
 The partition function with the Aharanov-Bohm flux $Z(\Phi)$
 has its minima for integer values of flux.
This feature is ascribed to the fact that
the $c=-1$ part of the permanent state has no background charge
\cite{FMS} while the $c =-2$ part of the HR state does.

The unitarity of the edge theory also is realized differently
from the HR state. In the HR state, the unitarity is restored
by the presence of spontaneous flux \cite{ino}.
This exchanges the couplings between the pairing and the charge
degrees of freedom, and the edge state becomes
identical to the 331 state with the $c=1$ chiral Weyl fermion as
its pairing degrees of freedom.
On the other hand, we don't have such spontaneous
flux in the permanent state.
However, it is known that the $c=-1$ theory generates the
Verma modules of the $c=2$ Virasoro algebra realized
on the $\Z_2$ orbifold of complex boson \cite{gurus}.
Thus the edge excitations of the permanent state can be
viewed as generated by the $c=2$ Virasoro algebra and
the unitarity is realized.
Note that it differs from the relation
between the $c=-2$ theory and the $c=1$ theory underlying the
HR state,  since the $c=-1$ theory only generates the
Verma modules of the $c=2$ Virasoro algebra
and  the correspondence is not extended to
 the one between the solitonic sectors of two theories.

The construction we have considered can be extended to
form hierarchy. This is possible since the permanent state
has an odd denominator filling fraction. A few examples
was given in \cite{YGM}, but no systematic construction
has been given. Actually this is readily done in the following way.
In general, the degrees of freedom of  $m$ filled Landau levels are
described by $m$ species of
chiral bosons
$\varphi_i  (i=1,\cdots,m).$  The pairing part of the
permanent state is expected to be
 unaffected by the formation  of the multi Landau levels
 of composite fermion.
The couplings between $\varphi_i$ are specified by
an integer matrix $K$ \cite{read,frohlich}.
The permanent analog of the Jain's hierarchy is
obtained by taking  $K$ as
\eqbegin
K_{ab} = \pm \delta_{ab} +  s C_{ab},
\eqend
where $s$ is a positive even integer and $C_{ab} =1 $ for
$\forall a,b =1, \cdots,m$.
The filling fraction is given by
\eqbegin
\nu= \frac{m}{ms \pm 1}.
\eqend
As shown in \cite{frohlich}, there is
 an affine SU($m$) symmetry of level $1$ in the $m$-th
 hierarchy.
The quasiholes form $m$ kinds of
 integrable representations of SU$(m)_1$,
which are labeled  as $a=1,\cdots,m$. They
correspond to $a$-th antisymmetric tensor of SU($m$)
respectively, which couples to charge $\frac{a}{m}\nu$
up to $Z\nu$.
Thus the primary fields for
the excitations on the permanent state are
$1V^{a}_{j} e^{i\frac{lm+a}{m\sqrt{q}}\varphi}$,
$\beta^{\alpha}V^{a}_{j}e^{\frac{lm+a}{m\sqrt{q}}\varphi}$,
$\sigma V^{a}_{j}e^{i\frac{lm+a}{2m\sqrt{q}}\varphi}$,
$\widetilde{I}V^{a}_{j}e^{i\frac{lm+a}{m\sqrt{q}}\varphi}$
$\beta\widetilde{I}V^{a}_{j}e^{i\frac{lm+a}{m\sqrt{q}}\varphi}$
where $l = 0,\cdots,(ms\pm1)-1$, $q=1/\nu$ and
$V^{a}_j$ are the primary fields for the representation $a$
respectively.
This completes our description of spin-singlet hierarchical
structure formed on the HR state.
We can similarly consider other hierarchical schemes which
involve SO($m$) or Sp($m$) affine symmetry.
It is interesting that there is also
 a Lorentzian affine symmetry by the presence of $\phi$.

We'd like to point out  that the Pfaffian or other candidate
states for the $\nu=5/2$ plateau cannot have such hierarchical
structure on them. It will simply violate the Fock condition
i.e. once we consider such a hierarchy,
the wave functions we  obtain cannot represent
the electron system.
Thus, if one assumes the validity of the idea of hierarchy
around $\nu=5/2$,  the existence of hierarchy on the HR state
 gives an evidence that  the
$\nu=5/2$ plateau is the HR state, and that
the odd-denominator plateaux around it ($ \nu = 7/3,8/3 ...$)
are the hierarchical permanent  QH states.

The major  experimental consequence of
our result is that  tilted field
experiments at these odd-denominator plateaux
 must behave similarly as the $\nu=5/2$ plateau
 \cite{eisen} i.e. as spin-singlet.
Unfortunately since, as far as we know,
the experimental works so far have been focussed
on the even-denominator FQH state,
we don't have information to confirm this at present.
Also the present experimental results are not
enough to determine which hierarchical scheme
is preferred around $\nu=5/2$, but
we expect that the permanent analog of the Jain's hierarchy we
constructed above may be most promising.

In this paper, we showed that the permanent
QH states are formed by the integer QHEs on the HR state.
Novel conformal field theory
(the extended $\beta$-$\gamma$ system)
description
was deduced from the modular invariance
 condition of the edge states for the state on the cylinder
 (see also \cite{subseq}).
We also developed a hierarchical scheme
of the permanent QH states, including the permanent analog of
 the Jain's hierarchy.
It yields a candidate  scheme for
understanding the FQHEs around $\nu=5/2$.
The major experimetal consequence is that
tilted field experiments
at the odd-denominator plateaux around $\nu=5/2$
 will show the spin-singlet behavior as
the enigmatic $\nu=5/2$ plateau. We  propose
 experiments to  test our prediction.

{\it Acknowledgement} The author would like to thank
M. Kohmoto and H.Kuriki  for useful discussions and M.Yamanaka
for comments on  the manuscript.

\vskip 0.2in
\noindent

\def\NP{{Nucl. Phys.\ }}
\def\PRL{{Phys. Rev. Lett.\ }}
\def\PL{{Phys. Lett.\ }}
\def\PR{{Phys. Rev.\ }}

\end{multicols}

\begin{thebibliography}{99}
\bibitem{willet}   R.L.Willet, J.P. Eisenstein, H.L. Stormer, D.C.Tsui,
A.C. Gossard and J.H. English, \PRL {\bf 59} (1987) 1776.
\bibitem{eisen}   J.P. Eisenstein et al., \PRL {\bf 61} (1988) 997;
 J.P. Eisenstein et al. Surf.Sci.{\bf 229}(1990)31.
\bibitem{halrez}  F.D.M. Haldane and E.H. Rezayi,  \PRL {\bf 60}
(1988) 956
: {\it ibid} 1886.
\bibitem{YGM2}   A.H. MacDonald, D.Yoshioka and S.M. Girvin,  \PR {\bf B39}
(1989) 1932.
\bibitem{morf} R.H.Morf, \PRL {\bf 80}(1998) 1505.
\bibitem{moore}   G. Moore and N. Read, \NP {\bf B360} (1991) 362.
\bibitem{jain}  J.K. Jain, \PRL {\bf63},(1989) 199; \PR {\bf
B40}(1989) 8079; \PR {\bf B41}(1990) 7653.
\bibitem{HLR} 	B.I.Haperin, P.A.Lee and N.Read, \PR {\bf B47}(1992) 7312.
\bibitem{YGM}  D. Yoshioka, A.H. MacDonald and S.M.Girvin,  \PR {\bf B38}
(1989) 3636.\bibitem{bonesteel}   N.E.Bonesteel, \PRL {\bf 82}(1999) 984.
\bibitem{FMS} D. Friedan, E. Martinec and S.Shenker, \NP {\bf B271}
(1986) 93 :
see also D. Friedan, S. Shenker and E. Martinec, \PL {\bf 160B}(1985) 55;
V.Knizhnik, \PL {\bf 160B}(1985) 403.

\bibitem{RR} N. Read and E.H. Rezayi, \PR {\bf B54}(1996) 16864.
\bibitem{gurus}		S. Guruswamy and A.W.W. Ludwig, \NP {\bf
B519}(1998) 661.
\bibitem{cappelli}     A. Cappelli and G.R. Zemba, \NP {\bf B490} (1997) 595.
\bibitem{milo} M.Milovanovi$\acute{{\rm c}}$ and N.Read, \PR {\bf B53}(1996)
13559.
\bibitem{gurarie}     V. Gurarie,  M. Flohr and C. Nayak,
\NP{\bf B498} (1997) 513.\bibitem{kausch} K.Kausch, {\it Curiosities at c= -2},
hep-th/9510149.
\bibitem{ino}     K. Ino, to appear in \PRL {\bf 82}(1999).
\bibitem{read} N.Read, \PRL  {\bf 65} (1990) 1502.
\bibitem{frohlich} J. Fr\"ohlich and A. Zee, \NP {\bf B364} (1 991)517.
\bibitem{subseq}
K. Ino, {\it On the conformal field theory of the permanent quantum Hall states}, 	in preparation.


\end{thebibliography}
\end{document}